# Intrinsically distinct hole and electron transport in conjugated polymers controlled by intra and intermolecular interactions

Giuseppina Pace,[1,*] Ilaria Bargigia,[1,2] Yong-Young Noh,[3] Carlos Silva,[2] Mario Caironi[1,*]

[1] Center for Nano Science and Technology@PoliMi, Istituto Italiano di Tecnologia, Via Pascoli 70/3, 20133 Milano, Italy
[2] School of Physics, School of Chemistry and Biochemistry, Georgia Institute of Technology, Atlanta, Georgia, 901 Atlantic Drive, Atlanta GA 30332-0400, USA
[3] Department of Chemical Engineering, Pohang University of Science and Technology, 77 Cheongam-Ro, Nam-Gu, Pohang 37673, Republic of Korea

*<u>Corresponding Author 1</u>: giuseppina.pace@iit.it
*<u>Corresponding Author 2</u>: mario.caironi@iit.it

**Abstract**

It is still a matter of controversy whether the relative difference in hole and electron transport in solution-processed organic semiconductors is either due to intrinsic properties linked to chemical and solid-state structure or to extrinsic factors, as device architecture. We here isolate the intrinsic factors affecting either electron or hole transport within the same film microstructure of a model copolymer semiconductor. Relatively, holes predominantly bleach inter-chain interactions with H-type electronic coupling character, while electrons' relaxation more strongly involves intra-chain interactions with J-type character. Holes and electrons mobility correlates with the presence of a charge transfer state, while their ratio is a function of the relative content of intra- and inter-molecular interactions. Such fundamental observation, revealing the specific role of the ground-state intra- and inter-molecular coupling in selectively assisting charge transport, allows predicting a more favorable hole or electron transport already from screening the polymer film ground state optical properties.

**Introduction**

Organic semiconductors have been extensively investigated in the past few decades, particularly in view of their applications in new lightweight and flexible microelectronics, optoelectronic and sensing devices.[1,2] Solution processable organic semiconductors have enabled the formulation of printable inks extending their application in large area and cost effective electronics.[3-5]

Recent progresses have led to the synthesis of a wide range of homopolymers and donor-acceptor (D-A) copolymers,[6] showing high charge carriers mobilities, in the range of 1-10 cm$^2$/Vs.[7] Among ambipolar co-polymers, diketopyrrole-pyrrole (DPP) derivatives[8-11] have shown field-effect mobilities consistently higher than 1 cm$^2$/Vs for both carriers.[12-14] High mobilities are favored by mesoscopically ordered packings favored by an improved backbone planarity,[15-17] higher molecular weight,[18] as well as side chains engineering[19,20]. It has been proposed that longer polymer chains can favor charge transport by enhancing the interconnectivity between ordered polymer phases and amorphous-like ones.[17] An important step forward to improving charge transport properties is the introduction of techniques that favor the unidirectional alignment of polymer chains.[21-23] There are consistent evidences showing that in organic field-effect transistors (OFETs) the current accumulated within the transistor channel strongly improves when polymer chains are anisotropically aligned perpendicularly to the source and drain electrodes, demonstrating the favorable charge transport along the polymer chain.[18,21,22,24]

In spite of the great advances made in the field, it is still a truly brain-teaser whether holes and electrons can be equally well transported within the same material microstructure once they are injected from the electrodes, or if there are extrinsic effects favoring either the one species or the other. Though interfacial phenomena can play a major role in unbalancing the charge transport of the two carriers, the tickling question regards the influence of polymer chain chemical composition and solid-state microstructure on a preferential hole or electron transport.

In their work, Bredas *et al.* investigated, from a theoretical point of view, the role played by parameters such as chemical composition, backbone planarity and extension of intramolecular and intermolecular π-π conjugation, in selectively affecting hole and electron hopping transport.[25] They particularly gathered attention on the strong impact played by the relative distance and orientation of neighboring chromophores, and on the superposition of their outer molecular orbitals. Very small distortion, of the order of a fraction of an Angstrom, can determine a wide variation of the chromophores' electronic coupling, having an impact on holes and/or electrons charge transfer probability. Such information is clearly embedded in the semiconductor ground state spectra at the solid state, but until now, no keys have been provided to obtain hole and/or electron mobility insights from such spectra, a clear limitation to their predictive power. Nevertheless, such predictability would be fundamental for identifying those design principles that can boost further the desired unipolar or ambipolar transport properties of new polymers.

Ideal candidates for addressing the above points and simultaneously studying the intrinsic nature of holes and electrons transport, are ambipolar polymer devices where both charges can be transported similarly well. Between polymers showing good balance between holes and electrons mobility, we selected a DPP derivative (diketopyrrolopyrrole-thieno[3,2-b]thiophene, DPPT-TT), where holes and electron transport can be investigated within the same device, while keeping the same chemical composition, thin film morphology and device architecture. Here, through Charge Modulation Spectroscopy and Microscopy (CMS and CMM) studies on operating organic field-effect transistors (OFETs),[26] as well as polarized UV-Vis and Transient Absorption (TA) spectroscopy acquired on thin DPP polymer films, we highlight for the first time the preferential ground state molecular interactions which can selectively favor either hole or electron transport in a donor-acceptor copolymer semiconductor. A first evidence of the different spectral signature of holes and electron polarons found in the charge absorption region was presented by Kathib *et al*.[27,28] However, the two polarons signature appeared as a broad band in the near-infrared (NIR) region making it difficult to provide information on their inter- or intra-chain nature. Here, by comparing polarization dependent ground state absorption spectra of anisotropically aligned polymer films with CMS bleaching spectra, we observe that holes and electrons selectively bleach different features of the ground state absorption. We found that specific inter- and intra-molecular interactions can favor either electron or holes transport, respectively.

The theoretical framework describing the electronic coupling between repeating chromophores in conjugated polymers that is at the base of such inter- and intramolecular interactions, was proposed by Prof. Spano.[29] Specifically, J-type electronic coupling occurs between covalently linked chromophores within the polymer chain, while H-type coupling involves the electronic coupling between cofacial chromophore units of neighboring polymer chains.

In this work, we show that the relative strength of intra-chain and inter-chain electronic coupling, identifying a more J- or H-type coupling respectively, sizes the ratio between electron and hole relative mobility. This correlation between carriers' mobility and spectral features makes it possible to predict from simple optical absorption measurements the balance between hole and electron transport. While no absolute mobility values can be extracted from absorption spectra due to the lack of a universal quantitative model, we show that the spectral variations observed on different polymers films, including the ones reported here and in previous literature, contain the necessary information for predicting the expected mobility trend in the investigate polymer films.

## RESULTS

### Aligned polymer films morphological and optical anisotropy

We selected a well-known DPP derivative, diketopyrrolopyrrole-thieno[3,2-b]thiophene (DPPT-TT).[30,31] As for other high molecular weight polymers owning thermotropic and/or lyotropic liquid crystal properties,[14,32] DPPT-TT can be easily aligned, allowing a high control on its film morphology, improved charge transport and electrical anisotropy.[18,24,31,33,34] In this work we employed the off-center spin coating technique to mechanically force the polymer chains alignment along the centrifugal force direction established during spinning.[34] In such films, DPPT-TT polymer chains have an edge-on backbone orientation and are aligned along the radial direction, as previously demonstrated.[11] When DPPT-TT polymer chains are aligned perpendicularly to the source and drain electrodes of a field-effect transistor, a ~ 40 times higher hole mobility can be measured with respect to the case where the chains are aligned parallel to the channel.[31]

Further evidence of the degree of polymer chains alignment is found in the angular dependence of the polarized absorption spectra (**Figure 1a**, **Supplementary Figure 1**). The maximum optical absorption occurs for an incident light polarized parallel to the chain alignment direction ($A_{//}$, parallel component of the absorption spectrum), while the minimum is observed when the light is polarized perpendicular to the chains ($A_\perp$, orthogonal component of the absorption spectrum). Two main peaks are present in both spectra, located at around 1.51 eV and 1.67 eV for $A_{//}$, and only slightly blue-shifted in $A_\perp$, similarly to what reported for DPPT-TT films aligned employing solution shearing[30] or slot die coatings.[12] For our films the dichroic ratio ($D$), defined as $A_{//}/A_\perp$, is ~ 2.6 at 1.51 eV (820 nm, **Figure 1a**).[35-37] The optical anisotropy data, along with previous 2D-GIXRD findings,[38] clearly indicate the presence of an electronic transition dipole moment (TDM) aligned along the polymer chain main axis.

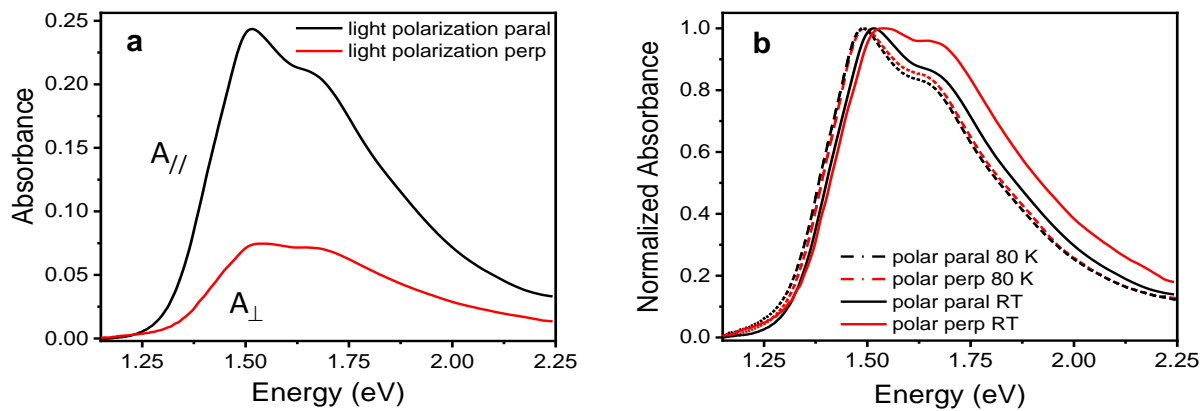

**Figure 1: UV-Vis and NIR absorption spectra.** a) Polarization dependent spectra of an off-center spin coated, aligned DPPT-TT film (from chlorobenzene solution, 10mg/ml); b) normalized data of the polarization dependent spectra at room temperature and at 80 K (trichlorobenzene solution, 10mg/ml). The temperature dependence is reversible upon cycling (**Supplementary Figure 2**). Light polarized parallel (polar paral, $A_{//}$) or perpendicular to the chain alignment direction (polar perp, $A_\perp$).

A weak shoulder located at ~1.45 eV, more pronounced in $A_\perp$, can be observed (**Supplementary Figures 3** and **Supplementary Note 1**). Since it is not possible to make any assignment of such red shoulder only on the basis of the absorption spectra, we further investigated its nature via Transient Absorption (TA) studies.

Due to their highest dichroic ratio (~ 0.59), TA spectra were acquired on films spun from chloronaphtalene (10 mg/ml) either exciting above the bandgap at 2.35 eV (**Figure 2a**) or at the red edge of the absorption spectra (pump at 1.42 eV, **Figure 2b and 2c**). The pump and probe light beams where polarized either parallel to the polymer chain main axis (pump/probe, paral/paral) or orthogonal (pump/probe, perp/perp). When exciting at high energy, the red shoulder at 1.42 eV appears more pronounced at longer time scale (~ 1 ps) due to its slower decay with respect to the rest of the ground state bleaching (**Supplementary Figure 4**). By exciting at the red edge of the absorption spectra (**Figure 2b**), we can observe that such excited state is directly accessible from the ground state. The time decay of TA acquired with paral/paral excitation shows a very fast initial component that dies off within 10 ps, followed by a very long lived dynamic (> 1ns; **Supplementary Figure 4**). These evidences strongly suggest that the red shoulder in the ground state absorption, appearing at ~ 1.45 eV and associated to a longer lived exciton, has a more marked charge transfer (CT) character. In the following, we will refer to such state as a CT exciton, in analogy to previous reports on similar features observed in other D-A low bandgap copolymers.[39-43] From the time decay of the TA spectra acquired with perp/perp excitations, it can be clearly observed that higher-energy peaks appearing at 1.51 eV and 1.67 eV are different in nature, being associated to shorter lived excited state (< 10 ps, **Supplementary Figure 4**).[44-46] A strong anisotropy of the TA spectra (paral/paral vs perp/perp) can be observed also at longer timescale (**Figure 2c**), where distinct relaxed excited states are present, demonstrating the very different energy landscape experienced by the excitons generated when the pump beam is either polarized parallel to the main chain axis or perpendicular to it.

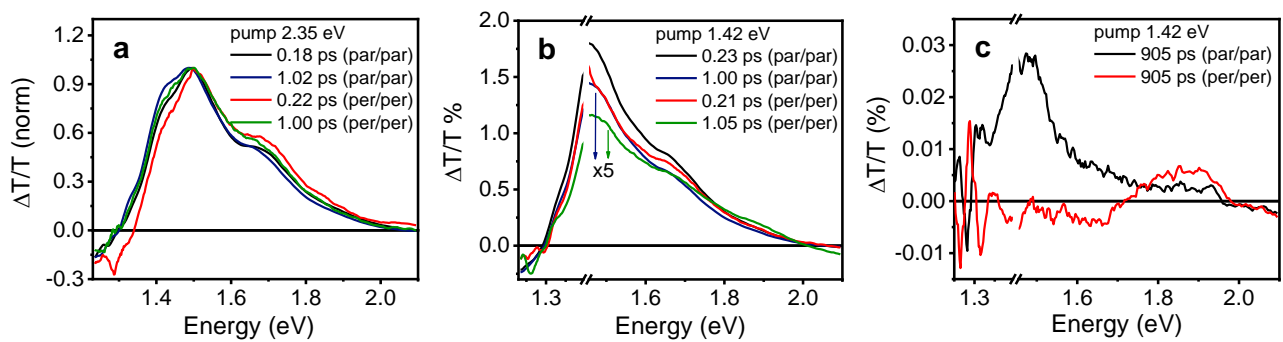

**Figure 2: Polarized transient absorption spectra (TA)**. Pump/probe light beams polarized either parallel (paral/paral) or perpendicular (perp/perp) to the polymer chain main axis. a) TA spectra acquired with a pump wavelength of 2.35 eV at 0.22 ps and 1 ps. b) TA spectra acquired with a pump wavelength of 1.42 eV at: b) 0.23 ps and 1.00 ps; c) 905 ps.

Therefore, we identified at least two distinct electronic transitions in the linear absorption spectrum, one at ~ 1.45 eV dominating the red shoulder that we assigned to a CT exciton, and the other characterizing the remaining visible region. The maximum TA signal found for the CT band is recorded when both pump and probe beams are polarized along the polymer backbone alignment direction, indicating that the transition dipole moment of the CT state ($TDM_{CT}$) lays parallel to the polymer chain (**Figure 2b**). The low energy feature at 1.51 eV of the ground state absorption is assigned to the 0-0 vibronic component of the $\pi-\pi^*$ transition and the 1.67 eV feature to the 0-1 one, in agreement with previous assignments in the literature for similar DPP derivatives.[47,48]

The vibronic features of the $\pi-\pi^*$ transition often observed in semiconducting polymers, can be mostly described within the framework of J- and H-type interactions.[49] J-type interactions identify intra-chain, through bond excitonic coupling, arising from a head-to-tail orientation of the chromophores placed along the polymer chain. H-type interactions refer to inter-chain excitonic coupling between face-to-face arranged chromophores belonging to neighboring polymer chains. The coexistence of H- and J-type interactions in disordered polymer films has been shown experimentally, and theoretically described,[50-52] for polydiacetylene (PDA)[52] and poly[2-methoxy-5-(2-ethylhexyloxy)-1,4-phenylenevinylene] (MEH-PPV)[51]. In conjugated molecular crystals,[42,53] two different ground states with orthogonal transition dipole moments having a distinct more prominent J or H character have been identified, and our observation well align with this conclusion which therefore extends also to highly ordered polymer films. H- and J-type interactions are characterized by distinctive signatures in the absorption and emission spectra and by a strongly different temperature dependence.[47-51] The relative strength of the progressing vibronic features of the π−π* optical transition, reveals which is the more prevalent interaction ($I_{0-0}/I_{0-1}$, where $I_{0-0}$ is the intensity of the 0-0 feature, and $I_{0-1}$ of the 0-1). This parameter is proportional to the Huang-Rhys factor ($\lambda^2$, $1/\lambda^2 \propto I_{0-0}/I_{0-1}$), which quantifies the coupling between an electronic transition and a phonon mode.[54] In particular, a decrease (increase) in the $I_{0-0}/I_{0-1}$ ratio, and a spectral blue (red) shift indicates the incurring presence of H- (J)-type interactions.[29,49] We observe that $I_{0-0}/I_{0-1}$ is equal to 1.19 in $A_{//}$, while it reduces to 1.08 in $A_{\perp}$ (**Supplementary Table 1**). As a reference, we extracted such ratio factor also in DPPT-TT solutions at low concentration, where single chains and mostly J-type interactions are present ($I_{0-0}/I_{0-1}$ = 1.24, **Supplementary Figure 5** and **Supplementary Table 1**). Therefore, J-type interactions are prevalent in $A_{//}$, reflecting their intra-chain nature along the polymer chain, while a stronger contribution of H-type interactions is present in $A_{\perp}$, as expected from their inter-chain nature. Evidences of the co-presence of two distinct H- and J-type optical transition dipoles with an almost orthogonal orientation have been previously found in small molecules crystals.[55,56] Such hybrid assignment of coexisting J- and H-type interactions within the same molecular packing and the presence of orthogonal excitons has not been studied to the same extent in polymer films, nevertheless there is a consistent number of reports identifying the coexistence of H- and J-type interactions also in this case.[50,55]

While photoluminescence (PL) is typically adopted to confirm the nature of the intra- and inter-molecular interactions, this is not viable for DPPT-TT, due to its extremely low PL quantum yield in solution, which becomes even lower in solid-state films.[10] We therefore resorted to temperature dependent absorption measurements. The temperature dependent spectral changes reported in **Figure 1b** are completely reversible (**Supplementary Figure 2**). At room temperature (RT), for $A_{\perp}$ we observe that $I_{0-0}/I_{0-1}$ = 1.06 (**Supplementary Table 1**). By lowering the temperature down to 80 K, the spectra show a ∼ 40 meV red shift of the $I_{0-0}$ peak and an increase in the $I_{0-0}/I_{0-1}$ relative peak intensity ($I_{0-0}/I_{0-1}$ = 1.20), indicating an increasing J-type character at lowering the temperature. Such H to J transition with temperature was previously observed in poly(3-hexylthiophene) (P3HT)[52,54] and poly(2-methoxy,5-(2'-ethyl-hexoxy)-p-phenylenevinylene) (MEHPPV),[36,51] implying that the overall photophysical response of such semiconducting polymers arises from the presence of both kind of interactions. The raising of the J-type feature at lower temperature can be rationalized in terms of reduced thermal disorder favoring a backbone planarization.[27,57]

**Charge-modulation investigation in field-effect transistors**

We here investigate in details the specific spectral features of holes and electrons transported in DPPT-TT OFETs, by means of CMS and CMM performed on devices where the polymer chains are aligned perpendicularly to the source and drain electrodes.[31] The ambipolar electrical characteristics are reported in the **Supplementary Figure 6**. CMS is a lock-in based technique that measures the variation in the optical transmission ($\Delta T$), normalized to the total transmission ($\Delta T/T$), of a semiconducting polymer film embedded in an operating OFET. The optical variation is caused by the frequency modulation of the accumulated charge carriers density due to the AC gate voltage superimposed to the DC bias. Only charges that are effectively contributing to the charge transport within the timeframe of the modulation frequency are probed, excluding deeply trapped, non-mobile states.[58,59] Therefore, CMS selectively probes those chromophores carrying a mobile charge in the few molecular layers lying below the dielectric interface where the charge accumulation takes place.[60-62] A positive CMS signal ($\Delta T/T > 0$), named bleaching, is a signature of an increased light transmission associated with a reduction of the density of absorbing neutral polymer segments, a consequence of the presence of mobile charges. Therefore, the bleaching region carries a direct spectral signature of the intra- and inter-molecular environment sensed by the charge. A negative CMS signal ($\Delta T/T < 0$) evidences the presence of new optical transitions generated by the absorption from the charged chromophores.

In **Figure 3** we report the CMS spectra acquired under hole and electron accumulation regimes, indicated in the following as *h-acc* and *e-acc*, respectively. Below 1.3 eV a broad charge absorption band appears in both spectra, while the 1.3 eV to 2 eV range is dominated by the ground state bleaching. **Figure 3a** shows the correspondence of the CMS bleaching peaks with the ones observed in the ground state absorption spectra. The finer structure found in the bleaching region for both *e*- and *h-acc* CMS spectra with respect to the UV-Vis absorption is a typical feature found in CMS spectra of different organic semiconductors, indicating that mobile charges preferentially populate regions of higher structural order.[58,59]

For both carriers, we distinguish three bleaching peaks appearing at 1.42 eV, 1.51 eV and 1.71 eV. The relative intensity of the 1.51 eV and 1.71 eV features, corresponding respectively to the 0-0 and 0-1 vibronic replica, is strongly different in the two cases. We further notice that the red shoulder at 1.42 eV, corresponding to the CT state, is much stronger in the *e-acc* spectrum than in the *h-acc* one, the latest being consistently blue shifted.

Very importantly, such observation of a strongly different relative intensities of the main bleaching peaks in CMS spectra under *h-acc* and *e-acc*, clearly indicates that holes and electrons selectively bleach different spectral components of the overall thin film ground state absorption.

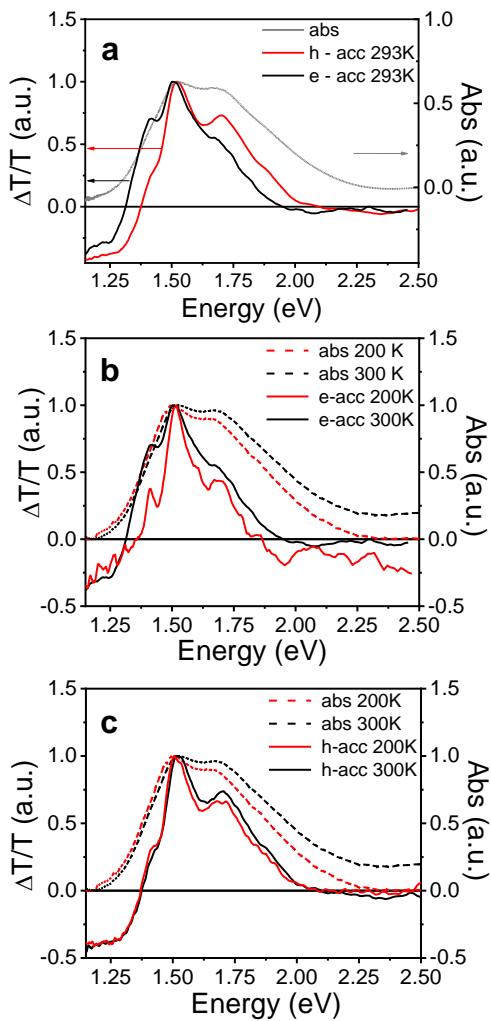

**Figure 3. Normalized temperature dependent CMS (continuous line) and absorption spectra (dashed line).** a) Comparison between absorption spectra, *h-acc* and *e-acc* CMS spectra acquired at room temperature; b) comparison between absorption spectra and *e-acc* CMS spectra acquired at 300 K and 200 K; c) comparison between absorption spectra and *h-acc* CMS spectra acquired at 300 K and 200 K (*e-acc*: $V_g$ = 30 V; $V_{pp}$ = 40 V ± 20 V; *h-acc*: $V_g$ = - 30 V; $V_{pp}$ = 40 V ± 20 V). The selected temperatures allow comparing the CMS spectra without incurring in the presence of lineshapes due to the interference from electroabsorption features (see **Supplementary Figures 7** and **8**).

By considering the relative intensity of the 1.51 eV and 1.71 eV features, we observe a stronger H-type character ($I_{0-0}/I_{0-1}$= 1.26) in the bleaching region under *h-acc*, while under *e-acc*, the bleaching reveals prevailing J-type coupling ($I_{0-0}/I_{0-1}$= 1.42). It is not possible to establish a well-defined cutoff value for such peak ratio that would enable to identify uniquely pure J-type interactions and pure H-type interactions since in polymer films they are mostly present concomitantly. However, a relative comparison will enable to follow the relative content variation in H- or J-type coupling. To corroborate this evidence, in **Figure 3b** and **3c** we overlapped the non-polarized ground state temperature dependent absorption spectra with the corresponding temperature dependent *e-acc* and *h-acc* CMS spectra. The temperature dependence of the CMS $I_{0-0}/I_{0-1}$ ratio confirms the increasing J character of the molecular interactions involved in the charge relaxation at lowering the temperature.[29,49,54,63] A weak red shift of the 0-0 peak from 1.52 eV to 1.50 eV in *h-acc* spectra can also be observed (**Figure 3c**), confirming the transition to increasing J-type character.[63]

The above experimental data represent the first fundamental evidence that holes and electrons relax within the polymer film selectively perturbing different intra and inter-molecular interactions: relatively, holes mostly bleach H-type inter-molecular interactions, while electrons preferentially bleach J-type intramolecular interactions. This fundamental finding highlights their intrinsic different transport characteristics.

Moreover, in the *e-acc* CMS the relative intensity of the 0-0 peak with respect to the CT peak strongly increases at lowering the temperature. Since the $TDM_{CT}$ state is aligned parallel to the polymer backbone, as evidenced by TA measurements, this might imply a stronger coupling between J-type interactions and CT states when an electron relaxes on the chain. At lowering the temperature, owing to the increased J-type interactions, a strong temperature dependence of the CT peak is indeed observed only in the *e-acc* CMS spectra, contrarily to the *h-acc* ones where the CT peak is already weak at room temperature.

By coupling CMS with an optical microscope, it is possible to spatially resolve, with a submicrometer resolution, the mobile charge signal within the OFET channel (charge modulation microscopy, CMM).[26,31,64-66] We also acquired local CMS spectra within the active channel, decoupling the CMS channel signal from eventual spurious effects, such as electroabsorption at the injecting electrodes.[67] The consistency of the spectral features found in the macroscopic CMS, acquired on the entire device, with the local CMS spectra demonstrates the absence of such interfering spurious effects (**Supplementary Figures 9** and **10**).

Light polarization dependent CMM allows to extract angular maps indicating the orientation of the optical TDM at a specific wavelength, and the degree of order (*DO*), quantifying the fraction of charge modulated signal which originates from anisotropically oriented TDMs (**Supplementary Note 2**).[65] **Figure 4** shows the angular and *DO* maps of the charge TDM ($TDM_{charge}$) acquired within the OFET channel under *e-acc* and *h-acc* regimes, respectively. At the chosen probing wavelength (1.18 eV), only the broad charge-induced absorption signal is detected, no ground state absorption exists and the anisotropic contribution to the angular map is associated only to the anisotropy of $TDM_{charge}$. We further confirm that for DPPT-TT transistors the $TDM_{charge}$ is aligned along the polymer chain main axis, providing a direct information on the preferential orientation of the conjugated segments probed by charge.[31,66] $TDM_{charge}$ angular maps found for holes and electrons are very similar, showing that in both cases $TDM_{charge}$ are predominantly aligned along the chains alignment direction (**Figure 4a, b**).

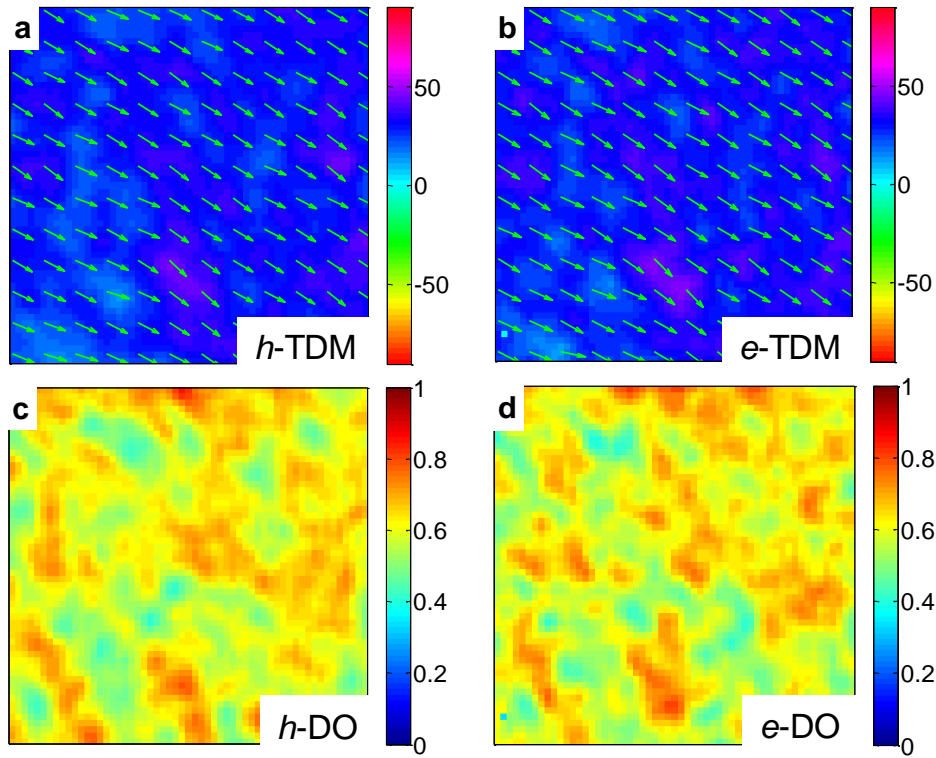

**Figure 4. CMM maps of the active channel probed at 1.18 eV, within the photoinduced charge absorption region.** (15μm x 15μm) Source and drain electrodes are located at the left and right side of the maps. a-b) Charge transition dipole moment (TDM$_{charge}$) maps acquired under *h-acc* and *e-acc*, showing the angular orientation of the anisotropic component of the charge modulation signal. Green arrows indicate that the TDM$_{holes}$ and TDM$_{electrons}$ mostly distribute along the polymer chains alignment direction. c-d) maps acquired under *h-acc* and *e-acc*, showing the local degree of order (*DO*), quantifying the weight of the anisotropic component of the charge modulation signal with respect to the isotropic component. Probing conditions: for *e-acc*, $V_g$ = 30 V, $V_{pp}$ = 40 V ± 20 V; for *h-acc*, $V_g$ = - 30 V; $V_{pp}$ = 40 V ± 20 V.

The recorded *DO* maps (**Figure 4c, d**) show that, in both accumulation regimes, there are regions with high degree of directional order (in the range from 0.7 to 1) widely present over the entire scanned area. Regions with *DO* ≈ 1 refer to areas where the entire charge modulation signal arises from very well aligned TDM$_{charge}$. Only in very limited regions, the TDM$_{charge}$ is more randomly distributed (*DO* < 0.2) (**Figure 5**). We found a perfect overlap of the CMS spectra collected in regions with *DO* ≈ 1 and *DO* < 0.2 for both accumulation regimes (**Figure 5a, b, c**). Such important evidence demonstrates that the nature of the charged state is not dependent upon the degree of the directional order of a specific region. The spectra found under *h-acc* and *e-acc* must therefore arise from the same ordered aggregate, which are present throughout the film independently from the local concentration of aligned chains and that are selectively probed by mobile charge. Therefore, the preferential bleaching of inter or intra-molecular interactions occurring during either electron or hole transport is uniquely selected by the nature of the mobile charge and not by the specific local morphology of the film.

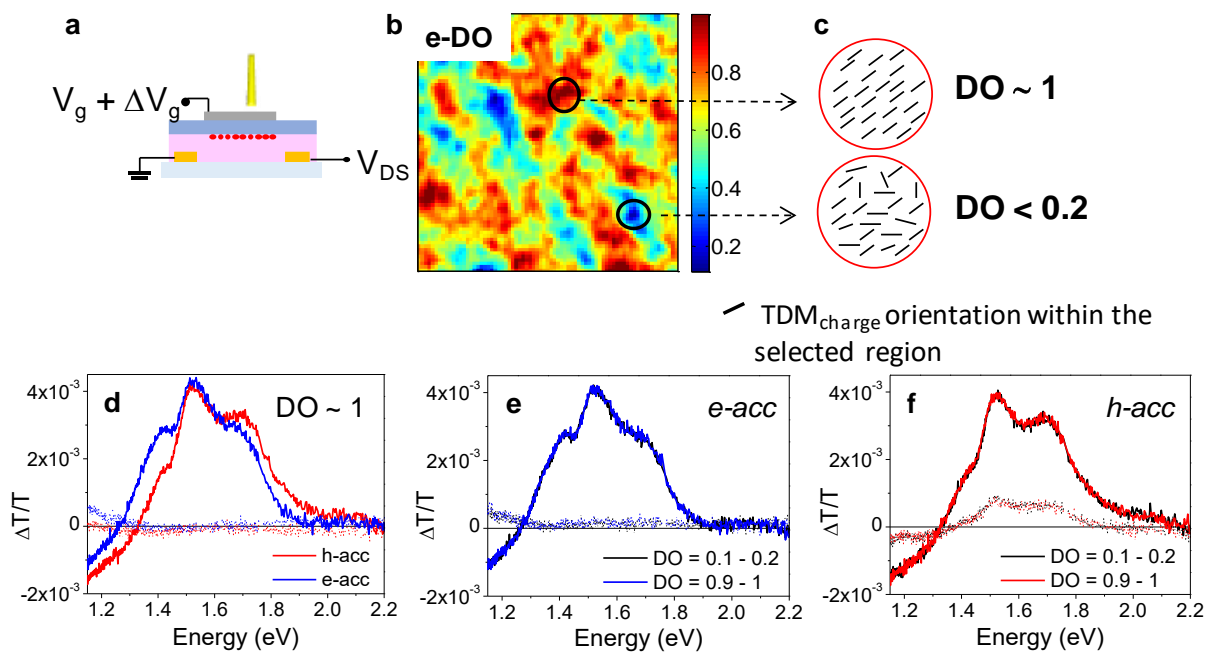

**Figure 5. Comparing local CMS spectra acquired in regions with *DO* ≈ 1 and *DO* < 0.2.** Top panel: a) scheme of the OFET device under local CMS light probe; b) *e-acc DO* map (15μm x 15μm); c) schematics of charge transition dipole moment alignment (TDM): black circles indicate regions of *DO* ≈ 1 and *DO* < 0.2, whose TDM vector distribution is sketched in the drawing on the right side; d) *h-acc* and *e-acc* spectra acquired in an area with *DO* ≈ 1; e) *e-acc* local CMS spectra acquired in regions with respectively *DO* ≈ 1 and *DO* ≈ 0.2; f) *h-acc* local CMS spectra acquired in region with respectively *DO* ≈ 1 and *DO* ≈ 0.2 (OFET electron-accumulation, *e-acc*: $V_g$ = 30 V; $V_{pp}$ = 40 V ± 20 V; hole-accumulation, *h-acc*: $V_g$ = - 30 V; $V_{pp}$ = 40 V ± 20 V; dashed lines, out of phase signal; degree of order, *DO*).

## Correlating transport properties to ground state transitions

We have so far demonstrated the different nature of the molecular interactions affected by holes and electrons when relaxing on a polymer site involved in transport. We now exploit such fundamental insight to show that it is possible to gain information on hole and electron transport already from the analysis of the ground state transitions. The lack of a generally applicable theory modeling organic semiconductors absorption bands, which could enable the precise extrapolation of the different overlapping vibronic peaks, does not allow to collect from those spectra absolute values that can be directly compared with mobilities values. Nevertheless, it is possible to take into account relative variations occurring along different polymer films absorption spectra and correlate those variations with mobilities changes.

We here correlate the solvent dependence of UV-Vis absorption spectra in DPPT-TT films, with their holes and electrons mobility. We recall that the degree of chain alignment, and the film dichroic ratio, achievable with off-center spin-coating is dependent on the solvent adopted in the DPPT-TT solutions, being highest in films spun from chloronaphtalene (cn), followed by films spun from trichlorobenzene (tcb), chloroform (cf), chlorobenzene (cb) and toluene (tol).[31] It was previously shown that holes mobility follows a similar trend, because of the positive effect of the degree of aligned chains on the charge mobility. We note here that also electrons mobility, despite being an order of magnitude lower, increases from toluene to chloronaphtalene in the same order as found for holes mobility (**Supplementary Figures 11** and **12**). Interestingly, from the UV-Vis absorption spectra it is possible to observe that also the CT contribution to the ground state absorption follows the same solvent dependence. Therefore, the highest mobility, both for holes and electrons, is reached in film spun from chloronapthalene, for which both the dichroic ratio and the ground CT state are higher than what found for the other solvents. The latter is in line with the observation of a cross coupling between the CT and the π-π* transitions found in the *e-acc* and *h-acc* CMS, suggesting that an increased content of aligned ordered aggregates in the films contribute to improve the mobility, through an increased uniaxial transport and CT state content.

Since the presence of either a hole or an electron along the polymer chain selectively bleaches H-type and J-type interactions, respectively, the question is if this different behavior of the charge surrounding can be interpreted in terms of a more or less favorable transport for each of those charges. In order to provide an answer to such question, we considered the relative ratio between electrons ($\mu_e$) and holes ($\mu_h$) mobility to assess the influence of those H- and J-type interactions on their relative transport.

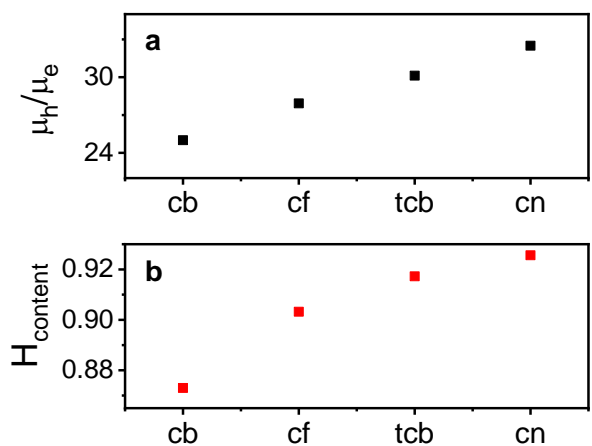

**Figure 6. Solvent dependence of mobility and $I_{0-0}/I_{0-1}$.** a) Solvent dependence of the holes and electrons mobility ratio ($\mu_e/\mu_h$). b) Solvent dependence of the relative peak ratio between the $I_{0-1}$ and $I_{0-0}$ peak ($\propto \lambda^2$) found in A$_{//}$ UV-Vis. Such ratio is proportional to the H-type electron-vibration coupling. Films spun from toluene are characterized by a negligible dichroic ratio and are not considered here. trichlorobenzene (tcb), chloroform (cf), chlorobenzene (cb), chloronapthalene (cn).

In **Figure 6** we compare the solvent dependence of the holes to electrons mobility ratio ($\mu_h/\mu_e$), with the H-type interactions content estimated from the $I_{0-0}/I_{0-1}$ value found in the $A_\perp$ of each polymer film showing optical dichroism. The two plots in **Figure 6** clearly show the steady increase of $\mu_h/\mu_e$ with the strength of H-type coupling within the aligned film. Therefore, not only holes are selectively bleaching H-type interactions, but the prevalence of H-type interactions at the ground state is shown here to be predictive of a more favored holes transport. At the same time, the presence of J-type interactions is predictive of a more favored electron transport in the polymer film.

Due to the rich amount of structural and charge transport investigations performed so far by different groups, it is possible to find ready available literature data to further confirm and generally extend the effectiveness of such ground state analysis for predicting the semiconductors charge transport properties, also in unipolar polymer devices.

In their work, Chen *et al.*[68] investigated the ambipolar properties of two polyselenophene derivatives. Relative prevalence of H- or J-type coupling signatures can be distinctly identified in both UV-Vis and photoluminescence spectra. From their data, an increasing H-type character observed in the ground state corresponds to a higher $\mu_h/\mu_e$ (**Table 1**), confirming our finding. In the same work, the authors also compare the same polyselenophene polymer backbone at varying the alkyl side chains group. Similarly to the previous observation, the different packing induced by the different side chains can promote either a relative increase of H- or J-type coupling content in the film. Also in this case, the correlation between H-type coupling content and holes mobility is present (**Table 1**). In a second work, Chen *et al.*[69] observed, in the case of a polyselenophene homopolymer, comparable CMS spectral features to the one found in this work for *h-acc* and *e-acc*, showing a stronger J-type character in *e-acc* CMS. Such observation is valid also when comparing mobilities values of different films of unipolar polymer devices. Interestingly, in the case of the n-type devices based on a well-known perylenediimmide co-polymer (P(NDI2OD-T2)) investigated by D'Innocenzo *et al.*,[70] the effect of polymer film annealing on the OFET performance could have been predicted from the analysis of the ground state spectra. The strong J-type character of P(NDI2OD-T2) is highlighted in its absorption spectra, coherently with its mostly favored electron transport properties in such device configuration. The electron mobility is higher for the pristine films, characterized by a stronger J-type character, than the melt-annealed ones (**Table 1**). Further examples showing the validity of our observation over different ambipolar and unipolar polymer devices, as well as homopolymers and D-A polymers, can be found briefly described in the **Supplementary Note 3**. Our findings are therefore solid with respect to the existing literature where sufficient data has been provided to test its validity.

Table 1: Relating relative mobilities with the relative content of H-type and J-type coupling in ambipolar polymer (P3OS,[69] PSS-C10, PSS-C8, PSS-C6,[68]) and electron transport to J-type coupling content in unipolar polymer (P(NDI2OD-T2))[70]. *Data from this work. $I_{0-1}/I_{0-0}$ is proportional to the Huang Rhys factor.

| Polymer Semiconductor | H-type coupling ($I_{0-1}/I_{0-0}$) | J-type coupling ($I_{0-0}/I_{0-1}$) | $\mu_h/\mu_e$ | $\mu_e$ (cm$^2$ V$^{-1}$ s$^{-1}$) | Reference |
|---|---|---|---|---|---|
| **P3OS** | 1.65 | | 5 | | [69] |
| **PSS-C10** | 1.14 | | 2 | | [68] |
| **PSS-C8** | 1.30 | | 3 | | [68] |
| **PSS-C6** | 1.75 | | 27 | | [68] |
| **P(NDI2OD-T2) pristine** | | 1.29 | | 0.056 | [70] |
| **P(NDI2OD-T2) melt annealed** | | 1.10 | | 0.041 | [70] |
| **DPPT-TT* (Chlorobenzene)** | 0.87 | | 25 | | |
| **DPPT-TT* (Chloroform)** | 0.90 | | 28 | | |
| **DPPT-TT* (Trichlorobenzene)** | 0.92 | | 30 | | |
| **DPPT-TT* (Chloronaphtalene)** | 0.93 | | 32 | | |

## DISCUSSION

We reported a series of coherent experimental observations that demonstrate intrinsic charge selectivity for intra- or inter-molecular interactions within the same semiconducting polymer film microstructure. The fine structure found in the CMS spectra confirms that both electrons and holes relax into ordered polymer aggregates. Both holes and electrons transport is favored by the alignment of the backbone chains and are maximized along the direction of the conjugated polymer backbone. The effectiveness of the solid-state film in transporting both holes and electrons can be monitored from the degree of alignment (dichroic ratio) extracted from a simple absorption spectrum. In D-A copolymers also the relative intensity of the CT state can be monitored to assess the increasing mobility by evaluating its relative weight in the ground state absorption spectra.

Most importantly, we showed that the transport of either a hole or an electron is selectively assisted by the presence of H-type interactions or J-type interactions, respectively. Though the absolute value of the charge mobility is principally related to the presence of a polymer ordered structure, chain alignment, presence of static and dynamic disorder, and ground state with CT character, the prevalence of either hole or electron transport is a function of the relative content of H- and J-type interactions.

For holes transport a strong perturbation of the H-type interactions occurs, as demonstrated by their optical bleaching. The correspondence between this selective perturbation and the increased hole mobility demonstrates that H-type intermolecular interactions mostly favor holes transport. Similarly, the increasing electron mobility with the increased J-type character of the ground state optical properties shows that J-type intramolecular interactions favor electrons transport.

The extensive consistency of our findings with previously reported data, collected both on D-A co-polymers and homopolymers, remark the general application of the proposed simple ground state analysis for the prediction of the field-effect device performances. We speculate that there must be a general reason for such specificity of holes preference to accommodate into H-type configurations and of electrons into J-type ones. A tentative explanation can be provided following first-order electrostatic arguments. At the first order, we can expect that the minimum relative distance to which cofacial chromophores can approach each other is limited by the increasing electrostatic repulsion between their outer electron densities. Therefore, in the presence of a negative charge, the best compromise for reducing such electrostatic repulsion is to relax into J-type intramolecular interactions rather than into H-type ones. The opposite occurs if a positive charge populates the aggregate, as an H-type configuration would lead a positive polaron to more favorably couple with the negative electronic cloud of the neighboring and cofacial chromophores.

Overall, with this work we propose the possibility of screening polymer solution formulations and film processing conditions towards optimal electron or hole transport properties, simply with ground state optical measurements, thus drastically reducing the effort and the cost of a full device fabrication and characterization through the typical "trial and error" approach. Furthermore, the charge selectivity upon specific molecular interactions can introduce a new parameter to be included within the design principles so far followed for achieving the desired unipolarity and ambipolarity of organic semiconductor devices. From a polymer design point of view, this might suggest new strategies to infer an H- or J-type character through the proper choice of heteroatoms, aromatic core dimension or side chains engineering. This study therefore unlocks a previously non investigated key aspect to boost even further the potential of organic semiconductors optoelectronics, fostering the development of analytical tools for drastically accelerating improvements in the field.

## METHODS

**Aligned film preparation.** The number-average molecular weight (Mn) of DPPT-TT is 70,000 g mol$^{-1}$ with a polydispersity index of 3.13, as determined by using high-temperature gel-permeation chromatography (GPC) at 135˚C with 1,2,4-trichlorobenzene as eluent. Solutions concentration was 10 mg/mL in chlorobenzene. For the off-center spin-coating, the clean corning glass substrate or the one carrying the electrodes pattern was located at approximately 2 cm from the spin-coater axis center (1,500 rpm, 3-second acceleration, in a $N_2$-purged glove box). The action of the centrifugal forces induced by the fast rotation of the spin-coater determines strong fluid-dynamic forces which induce the polymer chains to align along the radial direction. Not aligned films were obtained by placing the substrate at the center of the spin coater.

**Thin Film characterization.** UV−Vis spectrometry was conducted using a Cary Varian Eclipse spectrometer, and a polarized film was used in front of the beam source.

The DPPT-TT polymer films used for the ground state optical investigation were spin-coated from chlorobenzene solutions (10 mg/ml) at 1000 rpm.

**Temperature dependent optical transmission.** Samples were prepared through spin-coating from chlorobenzene solutions (10 mg/ml). Temperature dependent optical transmission spectra were acquired with a home-built Uv-Vis-NIR spectrometer with on a lock-in based technique. The set-up consisted of a tungsten lamp source, a monochromator, a Si-diode detector (Thorlabs FDS100), transimpedance amplifier (Femto DHPCA-100) and a lock-in amplifier (Standford Instrument SR830) connected to a computer. A matlab software was used to run the measurements. The transmission signal, was acquired by modulating the monochromatic light with an optical chopper (Stanford Instrument SR540). Typical transmission spectra were obtained dividing the modulated transmission acquired on the film ($T$) with the modulated transmission acquired on the reference glass sample ($T_0$). The absorption spectra was obtained from the its typical relation to the transmission signal ($-\log(T/T_0)$). A continuous-flow static exchange gas cryostat (Oxford Instruments) was employed with cryogenic liquid ($N_2$).

**Transient Absorption (TA).** Samples for TA studies were prepared on a glass substrate following the spin-off center coating procedure. A short pump pulse (∼220 fs, 530 nm or 855nm) was used as the photoexcitation pump beam (fluence always comprised between 8.6 μJ/cm$^2$ and 14 μJ/cm$^2$), while the excited states dynamics were probed with a delayed broadband probe pulse. Transmission changes $\Delta T$ were measured as a function of the pump–probe delay and of probe wavelength. The plotted signal is given by the differential transmission $\Delta T/T= [(T\text{pump on} - T\text{pump off})/T\text{pump off}]$. A mode-locked oscillator at 100 kHz (Pharos - Model PH1-20-0200-02-10, Light Conversion, Lithuania) emitting 1030nm pulses was used to generate both pump and probe beams. Pump wavelengths were generated coupling 10 W from the oscillator into a commercial optical parametric amplifier (Orpheus, Light Conversion, Lithuania), while probe wavelengths in the spectral range 500-1050 focusing 2 W of the oscillator power onto a sapphire crystal generated nm. Pump and probe pulses were focused and spatially overlapped on the sample, ensuring that the spot size of the probe beam was significantly smaller with respect to the pump beam. The transmitted probe signal was coupled into an imaging spectrograph (Shamrock 193i, Andor Technology Ltd., UK), combined with a multichannel detector. All measurements were taken with the samples in a vacuum chamber to prevent any influence from oxygen or sample degradation.

**Field-Effect Transistor Fabrication.** Prepatterned Corning Eagle 2000 glass substrates were cleaned by sequential ultrasonication in deionized water, acetone, and iso-propanol (15 min each). Source and drain electrodes were fabricated using conventional photolithography (channel width W and channel length ratio, W/L = 1.0 mm/20 μm), followed by thermal evaporation of Ni (5 nm) and Au (15 nm). The substrates were treated with oxygen plasma for 20 min before spin-coating the polymer solution. DPPT-TT solutions for devices were prepared with a concentration of 10 mg/mL in chlorobenzene. The off-center spin-coating was performed at 1500 rpm (3 s acceleration) in an $N_2$-purged glovebox. The electrode pattern was placed 20 mm away from the rotation axis of the spin-coater. The semiconductor layer thickness (30−50 nm) was measured with a surface profiler (Kosaka ET-3000i). The DPPT-TT film was thermally annealed at 200 °C for 20 min to remove the residual solvents under $N_2$ atmosphere.

Polymethylmethacrylate (PMMA, Aldrich, Mw = 120 kDa) was used as dielectric material without further purification, at a concentration of 80 mg/mL in n-butylacetate. The solution was filtered with a 0.45 μm PTFE syringe filter before spin-coating. After deposition, the devices were further annealed at 80 °C for 2 h under $N_2$. The optically semitransparent gate electrode was thermally evaporated and consisted of 4 nm Al top-gate electrode and 8 nm Au.

**Device Characterizations.** The OFET electrical characteristics were measured using a semiconductor parameter analyzer (Agilent B1500A) in an $N_2$-filled glovebox on a Wentworth Laboratories probe station. The field-effect mobility (μFET) was calculated from the saturation regime using equations for classical silicon MOSFETs as previously reported.

**Charge modulation spectroscopy.** All macroscopic CMS measurements are performed with a homemade vacuum chamber (~$10^{-5}$ mbar) in a transmission configuration. Source and drain electrodes were kept grounded, while the gate bias modulation was generated by a waveform generator (Keithley 3390) and a voltage amplifier (Falco Systems WMA-300). A tungsten lamp was employed as light source along with a monochromator. The light was focused on the device and the transmitted light signal was recorded with silicon photodetector (Thorlabs FDS100). The signal was then amplified with a transimpedance amplifier (Femto DHPCA-100) and then sent to a lock-in amplifier (Standford Instrument SR830) to obtain the differential transmission signal, $\Delta T$.

**Polarized Charge Modulation Microscopy.** The p-CMM data were collected with a homemade confocal microscope that operated in transmission mode. The light source consisted of a supercontinuum laser (NKT Photonics, SuperK Extreme) monochromated by an acousto-optic modulator (NKT Photonics, SuperK Select) in the 500−1000 nm region with line widths between 2 and 5 nm. Laser polarization was controlled with a half-wave plate and linear polarizer. The light was then focused on the sample with 0.7 N.A. objective (S Plan Fluor60x, Nikon) and collected by a second 0.75 N.A. objective (CFI Plan Apochromat VC 20x, Nikon). The collected light was focused on the entrance of a multimodal glass fiber with 50 μm core that acted as confocal aperture. Detection was operated through a silicon photodetector (FDS100, Thorlabs). The intensity of the collected light was registered by a silicon photodetector (Thorlabs FDS100). The signal was amplified by a transimpedance amplifier (DHPCA-100, Femto) and supplied both to DAQ (to record the transmission signal, T) and a lock-in amplifier (SR830 DSP, Stanford Research Systems) in order to retrieve the differential transmission data, $\Delta T$. The charge modulation signal is obtained by dividing the differential transmission signal, to transmission signal ($\Delta T/T$). A custom Labview program was used to run the measurements for data collection and to interface the laser source and a 3D piezo stage (Physik Instrumente P-517). The same software control of the piezoelectric element was used to perform local CMS spectra in a specific location within the OFET channel. Samples were placed in a homemade chamber with electrical feedthroughs to keep the sample under a continuous flow of nitrogen.

The voltage modulation at gate electrode is obtained by amplifying the voltage signal from the waveform generator (Keithley 3390) with a voltage amplifier (Falco Systems WMA-300). The gate modulation frequency was $f = 993$ Hz unless other specified. Prior to each measurement, the phase of the lock-in amplifier was synchronized to the amplified voltage modulation signal from the voltage amplifier at the first harmonic detection of the lock-in amplifier. A dual channel source-measure unit (Agilent B2912A) was used to keep the drain and source at ground.

**Data Availability**

Relevant data generated or analyzed during this study are included in this published article (and its supplementary information files). Any further source data are available from the corresponding author upon reasonable request.

**Acknowledgements**

We acknowledge Dr. Ajay Ram Srimath Kandada for discussions and suggestions on Transient Absorption experiments. M. C. acknowledges the support of the European Research Council (ERC) under the European Union's Horizon 2020 research and innovation programme 'HEROIC', grant agreement 638059. Y.-Y. N. acknowledges the support of the Center for Advanced Soft-Electronics (2013M3A6A5073183) funded by the Ministry of Science & ICT of Republic of Korea. The authors thank Ajay Ram Srimath Kandada for fruitful discussions. Authors are also thankful to Dr. Nam-Koo Kim (LG Electronics) for discussions.

**Authors Contributions**

P.G. and C.M. designed the experiments. P.G. and N.Y.Y. prepared and characterized the samples. I.B. acquired all TA measurements and performed the FC analysis. P.G. acquired CMS and CMM measurements and analyzed the data.
P.G. and C.M. wrote the paper. All authors discussed the data and edited the manuscript.

**Competing Interest**

The authors declare no competing interests.

# Supplementary Information

**Intrinsically distinct hole and electron transport in conjugated polymers controlled by intra and intermolecular interactions**

Giuseppina Pace,[1,*] Ilaria Bargigia,[1,2] Yong-Young Noh,[3] Carlos Silva,[2] Mario Caironi[1,*]

[1] Center for Nano Science and Technology@PoliMi, Istituto Italiano di Tecnologia, Via Pascoli 70/3, 20133 Milano, Italy
[2] School of Physics, Georgia Institute of Technology, Atlanta, Georgia, USA
[3] Department of Chemical Engineering, Pohang University of Science and Technology, 77 Cheongam-Ro, Nam-Gu, Pohang 37673, Republic of Korea

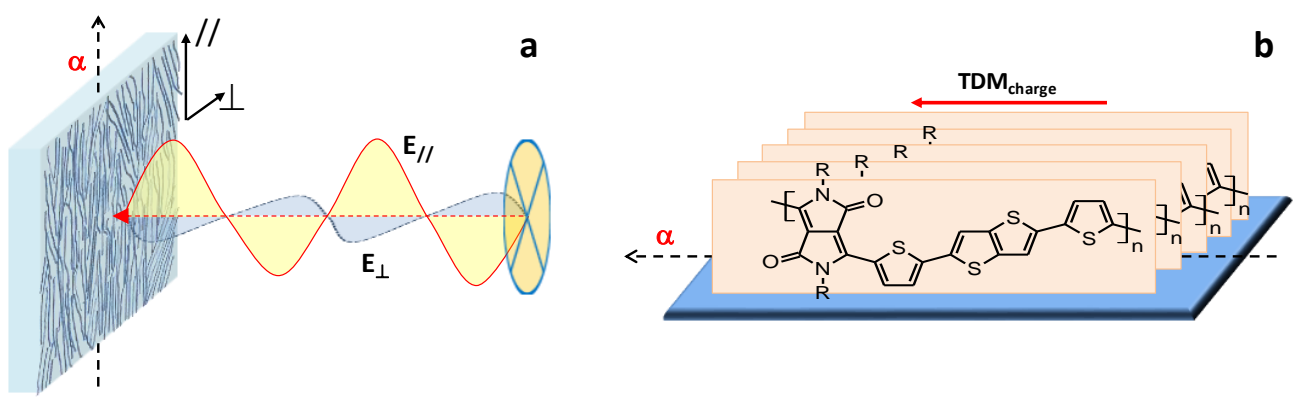

**Supplementary Figure 1: Sketch of an aligned polymer film.** a) Aligned polymer film under an incident polarized light beam whose electromagnetic field is either perpendicular ($E_\perp$) or parallel ($E_{//}$) to the polymer chain alignment direction ($\alpha$). b) Qualitative sketch of the edge-on configuration of the polymer chains at the top surface of the polymer film as found with 2D-GIXRD. The charge transition dipole moment ($TDM_{charge}$) orientation relative to the polymer chain alignment is also shown.

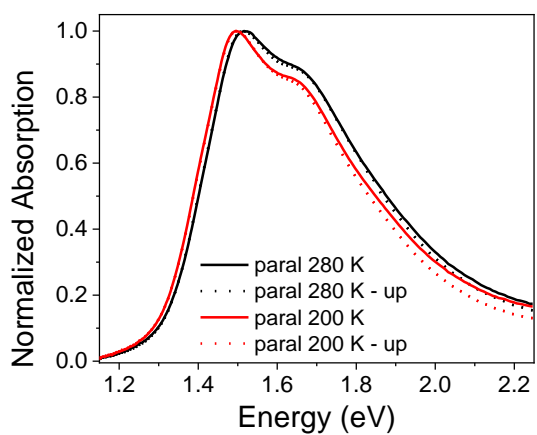

**Supplementary Figure 2: Temperature dependent UV-Vis.** The figure show the reversible absorption spectra acquired during the cycling of the temperature from 280 K down to 200K and up again to 200 K (up) and 280 K (up).

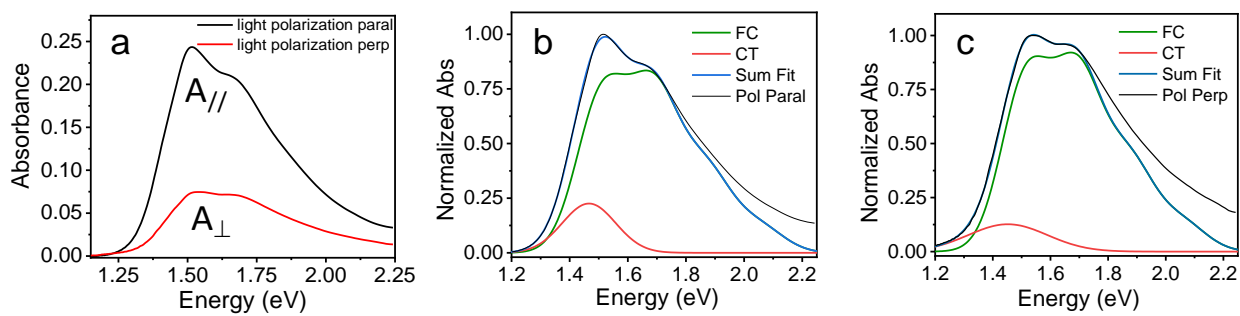

**Supplementary Figure 3: Franck-Condon analysis (FC) of spectra reported in Figure 1a of the main text.** a) Polarized UV-Vis absorption spectra: Light polarized parallel (polar paral, $A_{//}$) or perpendicular to the chain alignment direction (polar perp, $A_\perp$). FC of the Uv-Vis spectra acquired under incident parallel (b) and perpendicular polarized light (c). The Sum Fit spectrum is obtained from the sum of the FC curve fit of the main π-π optical transition and the CT (Gaussian) peak. A more prominent contribution of the CT peak is present in $A_{//}$ (b) in agreement with the TA findings.

**Supplementary Note 1. Franck-Condon analysis.** In **Supplementary Figure 3** we report a Franck-Condon analysis of the absorption spectra reported in Figure 1a of the main text. The Linear absorption data were fitted using a Frank Condon progression considering four vibronic levels and with the Huang Rhys factor, the energy of the 0-0 transition, the width of the gaussian, and a proportionality constant as free parameters. The energy of the C=C stretching mode was fixed at 0.18 eV. This analysis is not intended to be an accurate and rigorous study of the absorption peaks, which instead would require a more extensive theoretical work. The complexity of the system rely on the coexisting inter- and intrachain interaction and their spectral convolution with the CT state, whose contribution to the spectrum would require the difficult task to define a proper interaction Hamiltonian (Chem. Rev. 2018, 118, 7069−7163, Acc. Chem. Res. 2017, 50, 341−350). Nevertheless, we can demonstrate that the best fit of the spectra is obtained with a combination of a Franck-Condon (FC) fit for the π-π interaction and a Gaussian function for the red shoulder.

The FC progression perfectly fits the vibronic structure presents in the π-π transition, where a peak to peak distance of 0.18 eV is found between the vibronic peaks ($I_{00}$, $I_{01}$, $I_{02}$) and which falls into the vibrational region of the diketopyrrolepyrrole (DPP) unit C=C stretching mode (J. Mater. Chem. C, 2017, 5, 6176-6184). The red shoulder is instead spaced at 0.05-0.07 eV distance from the $I_{00}$ vibronic peak, showing that it does not belong to the same vibronic progression.

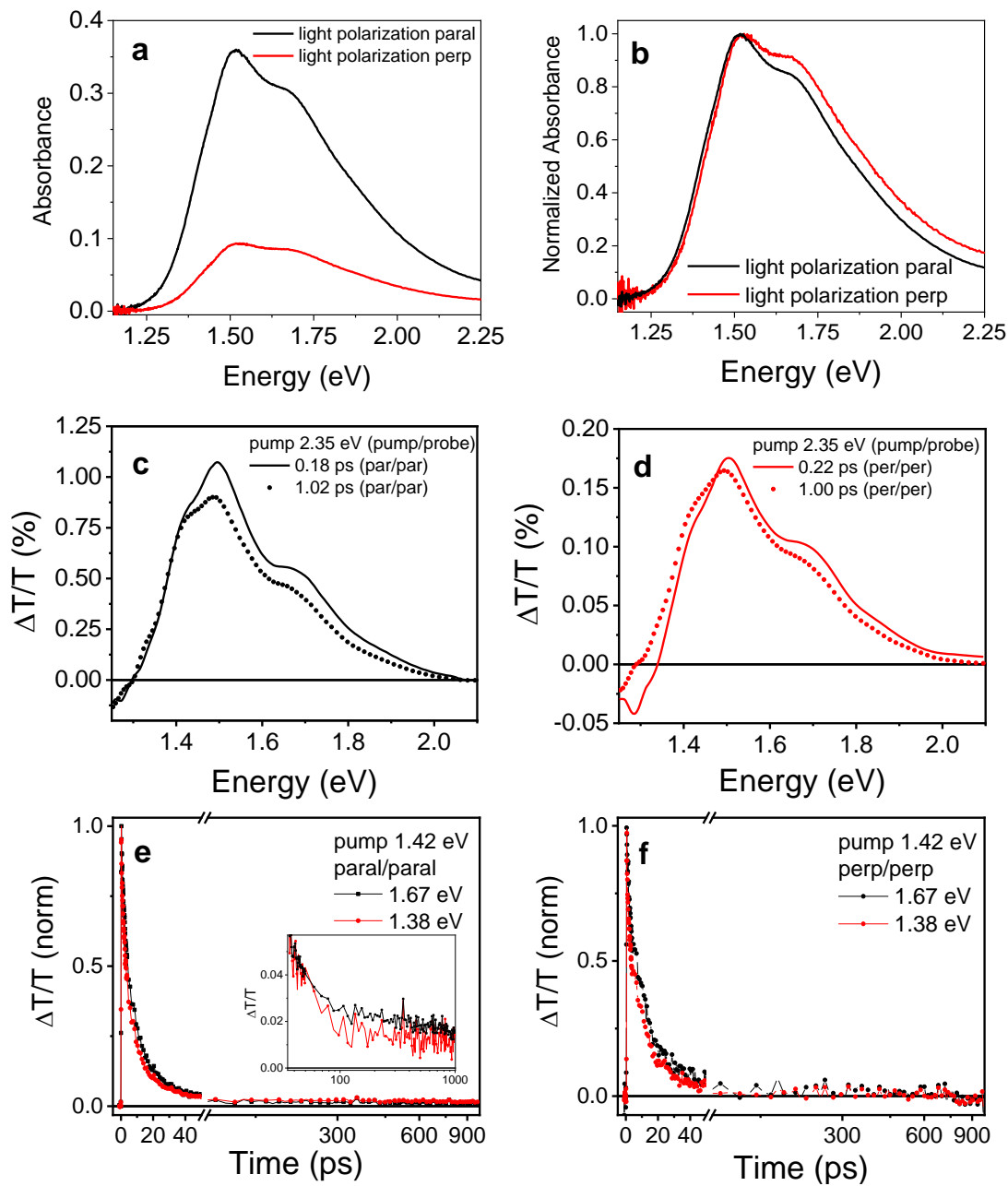

**Supplementary Figure 4: UV-Vis and Transient Absorption (TA) spectra.** a-b) Uv-Vis absorption spectra acquired under perpendicular and parallel incident light polarization; b- normalized spectra. c-d) TA spectra acquired at a pump beam 2.35 eV at varying the probe beam delay (~0.2 ps and ~ 1ps) showing the faster decay (~ 10 ps) characterizing the blue part of the spectra; e-f) Time decay of TA signal acquired at a pump beam of 1.42 eV with pump/probe polarization paral/paral (e) and perp/perp (f). The longer lived dynamic (up to 1 ns) which has been assigned to the CT ground state repopulation is clearly visible for pump and probe beams polarized parallel to the polymer chain (panel e), while a fast decay is present for the perp/perp (pump/probe, panel f). These data confirm the alignment of the CT transition dipole moment along the polymer chain. Film prepared with the spin-off center coating and from 10 mg/ml chloronapthalene solution followed by annealing at 200°C).

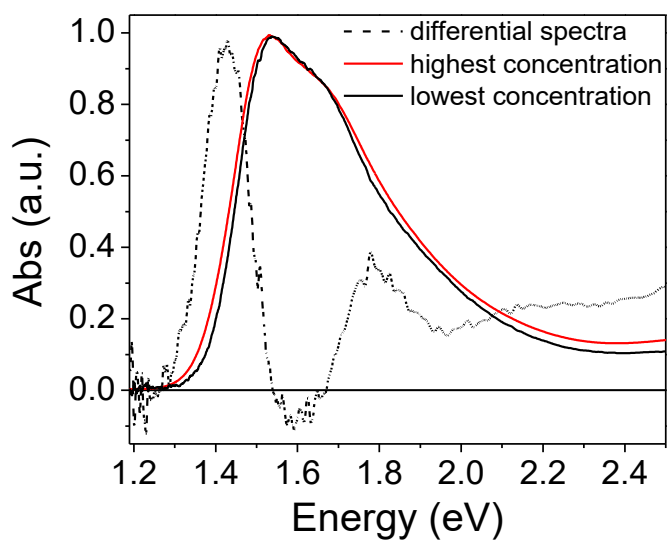

**Supplementary Figure 5: UV-Vis absorption spectra acquired in solution**. DPPT-TT solutions in chlorobenzene at the lower concentration (1.5 x 10$^{-6}$ M, red line) and at the higher concentration (17.5 x 10$^{-6}$ M, black line) showing the different absorption features of a non-aggregate vs an aggregated polymer chain. The dotted line shows the differential spectra highlighting the red shoulder prominence in the aggregate spectra appearing at 1.45 eV.

**Supplementary Table 1: $I_{0-0}/I_{1-0}$ ratio in solution and film.** Peak position associated to the absorption spectra acquired on DPPT-TT in solution and films. $I_{1-0}$ peak values were extracted from the second derivative of the absorption.

|  | $I_{0-0}$<br>eV (Intensity value) | $I_{1-0}$<br>eV (Intensity value) | $I_{0-0}/I_{1-0}$ |
|---|---|---|---|
| **Diluted solution** |  |  | 1.24 |
| *parallel component (fig 1a)* | 1.512 (1) | 1.658 (0.837) | 1.19 |
| *orthogonal component (fig 1a)* | 1.518 (1) | 1.651 (0.923) | 1.08 |
| *parallel component 80 K (fig 1b)* | 1.487 (1) | 1.658 (0.813) | 1.23 |
| *orthogonal component 80 K (fig 1b)* | 1.494 (1) | 1.662 (0.828) | 1.20 |
| *parallel component RT (fig 1b)* | 1.518 (1) | 1.685 (0837) | 1.19 |
| *orthogonal component RT (fig 1b)* | 1.535 (1) | 1.685 (0944) | 1.06 |
| **CMS *h-acc* local spectra** | 1.514 (4.3 x 10$^{-4}$) | 1.68 (3.4 x 10$^{-4}$) | 1.26 |
| **CMS *e-acc* local spectra** | 1.512 (4.4 x 10$^{-4}$) | 1.67 (3.1 x 10$^{-4}$) | 1.42 |

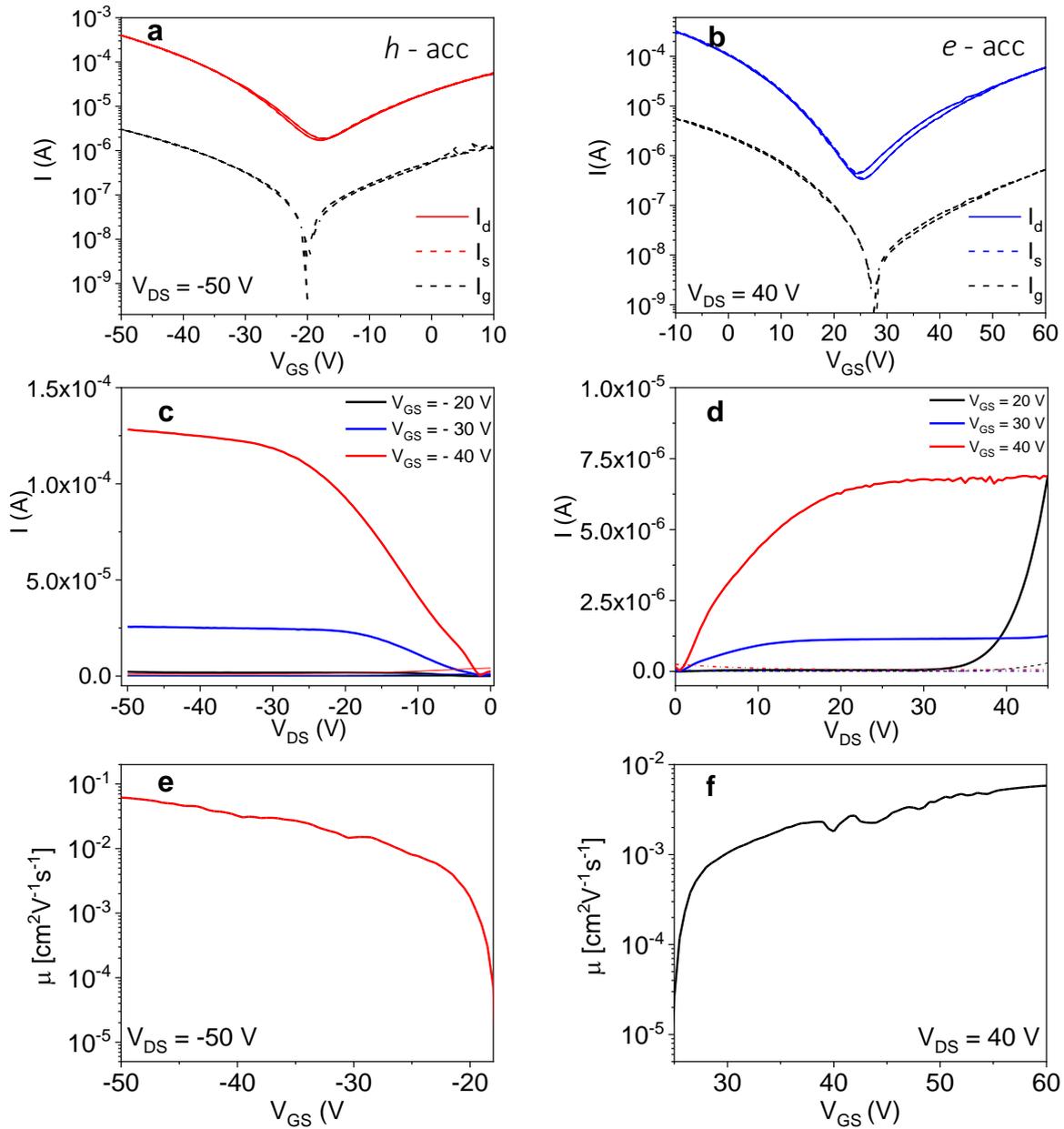

**Supplementary Figure 6: Electrical characteristics of DPPT-TT FET.** The polymer chains alignment is perpendicular to the channel length: a) and b) transfer curves; c) and d) output curves. e- holes and f- electrons mobility acquired in saturation regimes (channel width W, channel length ; W/L = 1.0 mm/20 μm; PMMA gate dielectric thickness 500 nm and capacitance ~6.2 nF/cm$^2$). The mobilities dependence from $V_{GS}$ in saturation regime, were calculated from the slope of $I_{drain}$ vs $V_{GS}$ ($\sqrt{I_{drain}}$ vs $V_{GS}$) according to the gradual channel approximation. For a more comprehensive study of the mobility dependence, please refer to Kim *et al.*[2]

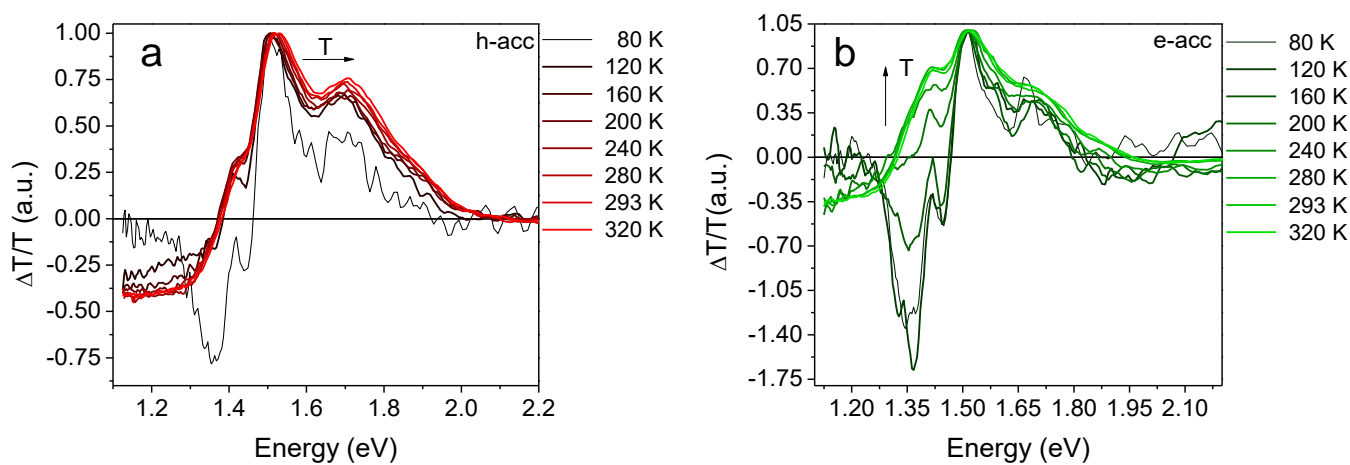

**Supplementary Figure 7: Normalized temperature dependent CMS.** Spectra acquired under holes accumulation a) and electrons accumulation (b) regimes. The contribution of electroabsorption starts being more visible at T lower than 150 K for e-acc regimes and at even lower T for h-acc. (*e-acc*: $V_g$ = + 30 V; $V_{pp}$ = 40 V ± 20 V); *h-acc*: $V_g$ = - 30 V; $V_{pp}$ = 40 V ± 20 V).

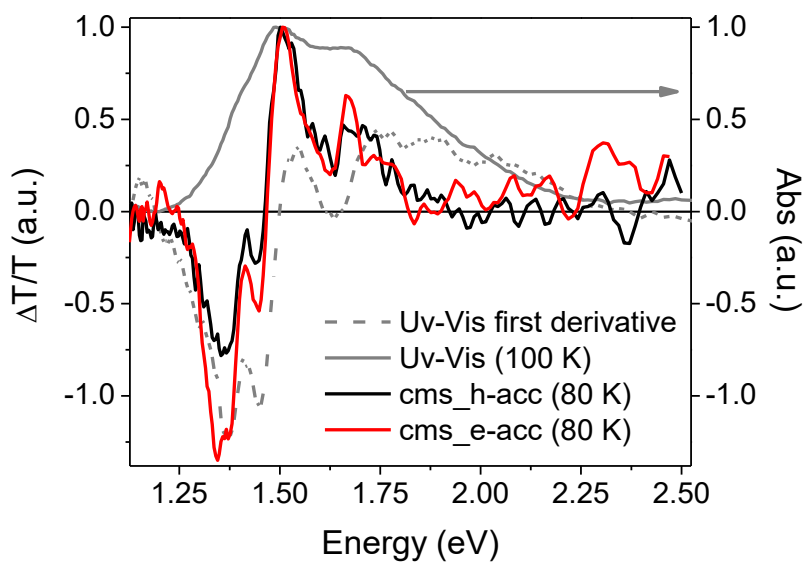

**Supplementary Figure 8: Comparing e-acc and h-acc spectra at low temperature (80 K).** In the absence of charge accumulation in the OFET the spectra resemble the first derivative of the J-like absorption spectra as expected. At low temperature (< 100 K) where no charge accumulation is occurring due to the increased injection barrier, the CMS spectra is mostly dominated by the electroabsorption (EA) features. We already observed that at low temperature, the planarization of the polymer backbone leads to an increase of J-like features, therefore, we observe the e-acc and h-acc spectra to converge to similar EA features.

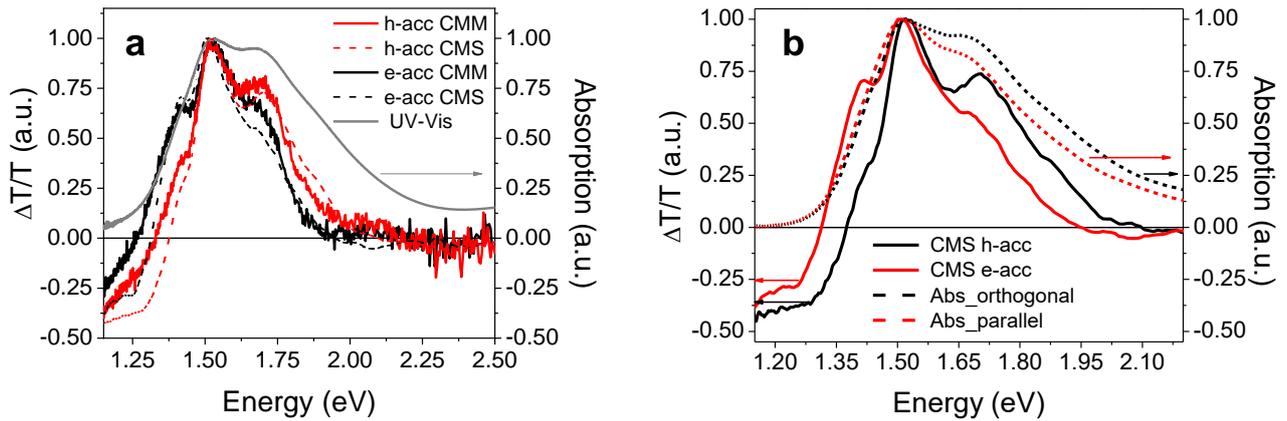

**Supplementary Figure 9: Macroscopic versus local CMS.** a- Comparison between charge modulation spectra acquired under h-acc and e-acc either locally within the active channels (CMM) or over the entire device area (CMS). b- Normalized room temperature CMS spectra overlapped with the parallel and orthogonal anisotropic components of the absorption spectra.

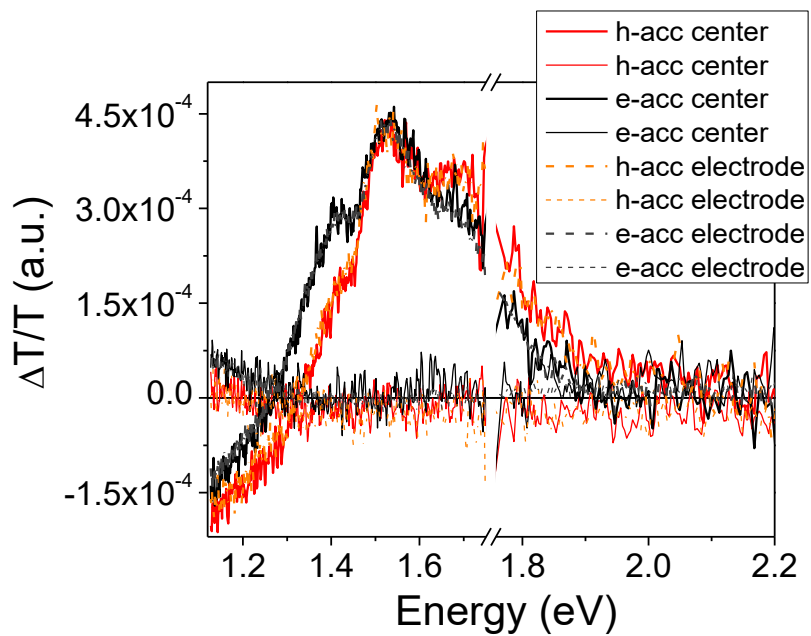

**Supplementary Figure 10**: **Local CMS within the FET channel and on electrodes.** No difference is observed in the local CMS spectra acquired either in the channel center, or at the electrode edge, implying the presence of low charge injection barrier at room temperature and the absence of spurious interferences arising from the gold electrodes. Out-off phase lock-in signal is also reported per each accumulation regime. (Gold electrodes are ~ 20 nm thick to enable optical transparency).

**Supplementary Note 2: Transition dipole moment ($TDM_{charge}$) and degree of order ($DO$).**

$TDM_{charge}$ ($TDM$) and $DO$ maps are acquired at six different laser polarization angles θ. The intensity of the charge modulation signal (CM) found per each image pixel coordinate (x,y) is defined as $I_{(x,y)}(\theta)$. Data are fitted according to the following equation[1]:

(1)
$$I_{(x,y)}(\theta) = M_{(x,y)} \cos^2(\beta_{(x,y)} - \theta) + C_{(x,y)}$$

where $C_{(x,y)}$ is the fraction of signal independent from the laser polarization and is related to the randomly distribute $TDM$. $M_{(x,y)}$ is the amplitude of the polarization dependent CM signal, and $\beta_{(x,y)}$ is the $TDM$ preferential alignment direction.

From the extracted $M_{(x,y)}$ and $C_{(x,y)}$ quantitative information on the fraction of aligned $TDM$ can be derived, according to the following equation:

(2)
$$DO_{(x,y)} = \frac{M_{(x,y)}}{M_{(x,y)} + 2C_{(x,y)}}$$

Factor 2 takes into account that anisotropic $TDM$ has halved probability, on average, to absorb a photon. The resulting $DO$ map indicates the fraction of CM signal arising from anisotropically distributed $TDM$.

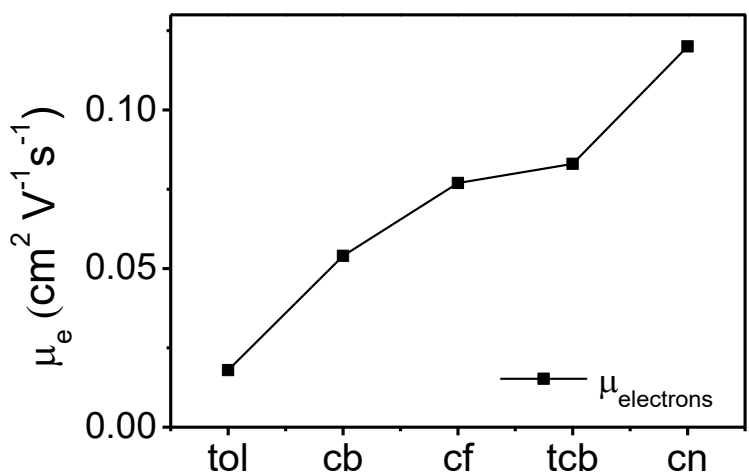

**Supplementary Figure 11. Solvent dependent electron mobility**. OFETs are prepared from off-center spin coated films. Polymer chains are aligned perpendicularly to the Source and Drain electrodes. The increasing mobility follows the increasing content of the CT ground state as found also for holes mobility.

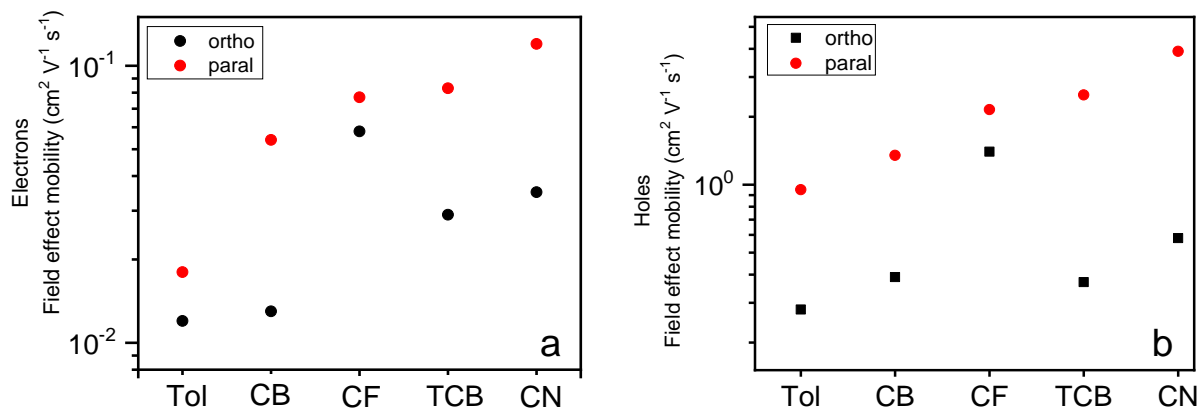

**Supplementary Figure 12: Comparing the solvent dependence of the electron (a) and hole (b) mobility.** Polymer chains in the devices were aligned either orthogonal or parallel to the source and drain electrodes. Data extracted as in reference 2 and 3.[2,3]

**Supplementary Note 3. Further examples of the mobility dependence upon J (H) character.**
The following examples are taken from the literature and support the main conclusion of this work.

*Adv. Mater.* **2018**, *30, 1704843*. In this paper a very pronounced J character can be observed in all the new polymers presented by the authors, and as expected all the reported polymers have a pronounced n-transport (e.g. N2200).

*Macromolecules*, **2006**, *39* (25), 8712. The authors present a series of conjugated thieno[3,4-b]pyrazine-based donor-acceptor copolymers at varying acceptor and/or side groups. The absorption spectra reported in Figure 3 of the cited paper show polymers with more pronounced H-character that well aligning with the observed hole transport for all polymers. Furthermore, a better resolved H-type vibronic structure is observed for the BTTP-F polymer followed by the BTTP-P and BTTP one. Table 1 reported by the authors shows how the highest hole mobility is indeed found for the BTTP-F polymer.

**Supplementary References**